%% file: main.tex
\documentclass[sigconf,natbib=false]{acmart}
\AtBeginDocument{%
  }

\usepackage{algorithm}
\usepackage{algorithmic}
\usepackage{makecell}
\usepackage{multirow} 
\usepackage{booktabs}
\usepackage{ragged2e}
\usepackage{graphicx}
\usepackage{balance}
\usepackage{textcomp}
\usepackage{xspace}
\usepackage{xcolor}

\usepackage{url}
\usepackage{colortbl}
\usepackage{hhline}
\usepackage{relsize}
\usepackage{pifont}
\usepackage{float}
\usepackage{hyperref}

\usepackage{circledsteps}
\usepackage{siunitx}
\usepackage[font=small]{caption}
\usepackage{academicons}

\hypersetup{
    colorlinks=true,
    linkcolor=blue,
    citecolor=blue,
    urlcolor=blue
}

\RequirePackage[
  datamodel=acmdatamodel,
  style=acmnumeric,
  sorting=none,
]{biblatex}
\AtEveryCite{\color{blue}}

\newcommand{\cmark}{\textnormal{\textcolor{black}{\ding{51}}}}
\newcommand{\xmark}{\textnormal{\textcolor{gray}{\ding{55}}}}
\definecolor{orcidlogocol}{HTML}{A6CE39}

\newcommand{\Design}{\texttt{\textbf{XL-HD}}}
\newcommand{\HDesign}{\texttt{\textbf{H-XL-HD}}}

\renewcommand\footnotetextcopyrightpermission[1]{}
\begin{document}

\title{\Design: Extended Learning in Hyperdimensional Computing via Deterministic Projections for In-Memory Accelerators}

\pagestyle{plain}
\fancyhead{}

\author{Sabrina Hassan Moon}
\affiliation{%
  \institution{\small University of South Florida}
  \city{ \small Tampa, FL}
  \country{\small USA}}
\email{ms38@usf.edu}

\author{Abu Kaisar Mohammad Masum}

\affiliation{%
  \institution{\small University of Louisiana at Lafayette}
  \city{\small Lafayette, LA}
  \country{\small USA}}
\email{c00591145@louisiana.edu}

\author{Sercan Aygun}
\affiliation{%
  \institution{\small University of Louisiana at Lafayette}
  \city{\small Lafayette, LA}
  \country{\small USA}}
\email{sercan.aygun@louisiana.edu}

\author{Dayane Reis}
\affiliation{%
  \institution{\small University of South Florida}
  \city{\small Tampa, FL}
  \country{\small USA}}
\email{dayane3@usf.edu}

\renewcommand{\shortauthors}{Moon et al.}

\begin{abstract}
Hyperdimensional computing (HDC) is a promising approach for energy-efficient edge machine learning (ML), where low latency, low power, and tight memory budgets are essential. However, traditional HDC relies on symbolic binding and pseudo-random high-dimensional vectors, which require large dimensionality and heuristic updates to reach competitive accuracy, limiting deployment on edge hardware.
We introduce~\Design, a deterministic, projection-based, fully learnable HDC framework tailored for in-memory acceleration within edge computing systems. The method uses a fixed Sobol sequence to project binary inputs, extending learning beyond conventional HDC. During training, class prototypes are optimized in real-valued space and later binarized, enabling an entirely binary dot-product inference pipeline ideal for IMC hardware such as ReRAM crossbars. \Design~achieves competitive accuracy on MNIST, UCIHAR, and ISOLET while maintaining a compact IMC-based inference engine with $0.395 \ \text{mm}^2$ area and only $0.40 \ \mu\text{J}$ per single-cycle inference. 
\end{abstract}


\maketitle
\input{01_Introduction}

\input{02_Background}
\input{03_Methodology}

\input{04_Hardware}

\input{05_Experimental_Results}

\input{06_Conclusion}
\input{07_Ack}

\printbibliography

% \appendix
\end{document}

%% file: 01_Introduction.tex
\section{Introduction}

With edge devices rapidly proliferating, machine learning (ML) frameworks require lightweight, fast, and energy-efficient processing to satisfy strict latency, bandwidth, and power limits. However, deploying conventional deep learning models remains challenging due to their high computational and memory demands, \textcolor{black}{motivating the use of in-memory computing (IMC) for massively parallel, energy-efficient processing directly within memory arrays~\cite{roy2020memory}.}

Hyperdimensional computing (HDC) has emerged as a promising alternative to computationally-intensive ML algorithms~\cite{kleykoSurvey, aygun2023learning}.
% Inspired by brain-like computation, 
HDC encodes scalar and symbolic data using high-dimensional vectors called hypervectors (HVs). However, generating these HVs -- either on the fly or by fetching them from memory -- incurs substantial energy and latency costs, especially in resource-limited environments~\cite{Masum2025HadamardHDC}. Improving vector generation and memory access, therefore, directly enhances baseline HDC performance. Once generated, simple operations (e.g., logic \texttt{XOR}, shifting, and population count (\texttt{POP+})) complete the encoding~\cite{rahimi2016robust}. The resulting HVs are naturally noise-robust, fault-tolerant, and effective in low-precision settings, making HDC well suited for edge AI.

\begin{figure}[t]
    \centering
    \includegraphics[width=\linewidth]{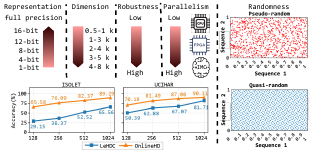}
    \vspace{-15pt}
    \caption{Trade-offs in bit precision, dimensionality, robustness, and parallelism (vs. OnlineHD~\cite{hernandez2021onlinehd} and LeHDC~\cite{duan2022lehdc}).}
    \label{fig:fig1}
    \Description{A figure showing trade-offs in bit precision, dimensionality, robustness, and parallelism for different HDC methods.}
   
    \vspace{-15pt}
\end{figure}

\textcolor{black}{Despite its potential, HDC faces key limitations for practical edge deployment, as summarized in Fig.~\ref{fig:fig1}. The top row highlights the core trade-off: when moving from full-precision to 1-bit encodings, conventional HDC must greatly increase dimensionality (D, from $0.5$–$1$K to $4$–$8$K) to maintain accuracy. The bottom plots confirm this trend, with LeHDC~\cite{duan2022lehdc} and OnlineHD~\cite{hernandez2021onlinehd} showing accuracy drops for D$ \leq 1$K. Since storage and compute scale with D, this dimensionality inflation directly raises hardware cost.
The platform comparison in Fig.~\ref{fig:fig1} shows that IMC offers the highest parallelism, making it ideal for low-precision HDC. A second challenge is the use of pseudo-random HVs, which scatter irregularly in the vector space and require large D for separability. In contrast, quasi-random sequences distribute points more uniformly, enabling comparable discriminability at much lower dimensionality. Together, these observations motivate our deterministic, hardware-efficient design that preserves HDC’s robustness and 1-bit parallelism while avoiding the high-D overhead of conventional methods.}

In this work, we present~\Design \footnote{ 
\href{https://github.com/ACEDLab/XL-HD}{
\raisebox{-0.30em}{\includegraphics[height=1.8em]{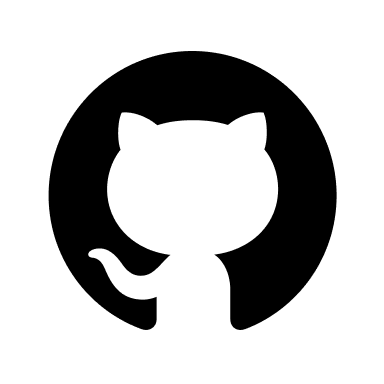}}%
}\textcolor{blue} {\textit{https://github.com/ACEDLab/XL-HD}}}, an algorithm-hardware \ co-designed HDC framework with extended learning capabilities and hardware-efficient edge inference tailored for in-memory accelerators, such as Resistive RAM (ReRAM) crossbars, offering substantial gains in speed, area, and energy efficiency. 

The main contributions of this work are as follows: (1) We propose~\Design, a deterministic HDC pipeline that uses Sobol-based binary mappings to streamline encoding and eliminate the overhead of traditional symbolic binding; (2) we introduce in-situ binarization of the encoded query HV and design ReRAM-based IMC hardware that performs inference entirely within crossbar arrays; and (3) we realize a complete end-to-end in-memory inference flow on ReRAM crossbars, achieving single-cycle throughput, sub-$\mu J$ inference energy, and a compact $0.395 \ mm^2$ area.

%% file: 02_Background.tex
\section{Background and Related Work}
\label{background}
\begin{table}[t]
\centering
% \color{purple}
\caption{Comparison of existing HDC and~\Design~accelerators.}
\label{tab:related_comparison}
\vspace{-10pt}
\scriptsize
\setlength{\tabcolsep}{0.96pt}
\renewcommand{\arraystretch}{0.96}

\resizebox{\columnwidth}{!}{%
\begin{tabular}{lcccccccc}
\toprule
\textbf{Approach} &
\multicolumn{3}{c}{\textbf{Encoding Method}} &
\multicolumn{2}{c}{\textbf{Efficiency}} &
\multicolumn{2}{c}{\textbf{Hardware Mapping}} &
\textbf{E2E}\\

\cmidrule(lr){2-4}
\cmidrule(lr){5-6}
\cmidrule(lr){7-8}
\cmidrule(lr){9-9}

 & \textbf{Pseudo} & \textbf{Symbolic} & \textbf{Deterministic}
 & \textbf{Low-D} & \textbf{Binary}
 & \textbf{IMC} & \textbf{ReRAM}
 & \textbf{IMC}\\

 & \textbf{Random} & \textbf{Binding} & \textbf{LD}
 & \textbf{Encoding} & \textbf{Inference}
 & \textbf{Support} & \textbf{Crossbar}
 & \textbf{Pipeline}\\

\midrule

OnlineHD~\cite{hernandez2021onlinehd}
& \cmark & \cmark & \xmark & \xmark & \cmark & \xmark & \xmark & \xmark \\

LeHDC~\cite{duan2022lehdc}
& \cmark & \cmark & \xmark & \xmark & \cmark & \xmark & \xmark & \xmark \\

FSL-HD~\cite{xu2023fsl}
& \cmark & \cmark & \xmark & \xmark & \xmark & \cmark & \cmark & \xmark \\

ReHDC~\cite{liu2024reram}
& \cmark & \cmark & \xmark & \xmark & \xmark & \cmark & \cmark & \xmark \\

ReX-HD~\cite{moon2025rex}
& \cmark & \cmark & \cmark & \xmark & \xmark & \cmark & \cmark & \xmark \\

\midrule
\rowcolor{gray!10}
\textbf{Ours (\Design)}
& \xmark & \xmark & \cmark & \cmark & \cmark & \cmark & \cmark & \cmark \\

\bottomrule
\end{tabular}%
}
\vspace{-10pt}
\end{table}

In this section, we provide an overview of the principles of HDC and IMC and discuss prior works that integrate HDC into IMC. 
% \vspace{-5pt}

\subsection{HDC Principles and Properties}

\textcolor{black}{HDC represents each data point (\textit{symbolic}, \textit{scalar}, or \textit{temporal}) by nearly orthogonal HVs that consist of hundreds or even thousands of binary (0,1) or bipolar ($\pm 1$) elements. To generate the symbolic HVs, a random number in $[0, 1]$ is drawn for each of the D dimensions and compared to $0.5$, producing a binary vector with roughly half 1s and half 0s. In contrast, scalar values are encoded by comparing the scalar to the same set of random numbers, resulting in correlated HVs for similar scalar values and dissimilar HVs for distant ones. Thus, random numbers play a central role in encoding both symbolic and scalar information, and the quality and characteristics of the randomness source are equally critical~\cite{thomas2021theoretical}. }

\textcolor{black}{ After generating HVs, HDC encoding applies the key operations: \ding{172} \textit{binding}, \ding{173} \textit{bundling}, and \ding{174} \textit{shifting}, depending on the application 
%(e.g., image processing utilizes operations \ding{172} and \ding{173}, while text processing needs \ding{174} as well)
~\cite{Kazemi2022}. \textit{Binding} combines two HVs into a new composite HV; using element-wise multiplication or \texttt{XOR} operation
for bipolar or binary vectors, respectively. \textit{Bundling} merges multiple HVs into a single robust and holistic representation. \textit{Shifting} maintains orthogonality and positional information of symbols by bit-shuffling the HV~\cite{rahimi2016robust}.
% (e.g., character order in text). 
%Traditional HDC employs \textit{pseudo-random} methods to generate HVs, which produce nearly orthogonal vectors but lack the structural properties required to achieve precise correlations. Low-discrepancy (LD) sequences, such as Sobol, address this limitation by distributing points more uniformly throughout the space, enabling the system to generate dependable correlated or uncorrelated HVs, even when using significantly smaller D~\cite{Mehran-No-Mult2023}.
Traditional HDC typically generates HVs using pseudo-random sources, which produce near-orthogonal vectors due to random fluctuations and offer limited control over the similarity structure, requiring a larger D to average out the randomness. In contrast, low-discrepancy (LD) sequences such as Sobol are deterministic mathematical constructions that distribute samples more uniformly and recurrently over the space, avoiding repeated random process while providing regular and reproducible deterministic HV patterns. This property can preserve accuracy even at significantly smaller D~\cite{Mehran-No-Mult2023}.
}
% \vspace{-5pt}
\subsection{IMC Primitives and Application in HDC}

% The growing demand for low-power, high-throughput edge computing has made the IMC a leading architecture. 
% Performing computation directly within the memory arrays, IMC addresses the von-Neumann bottleneck, enabling energy-efficient and massively parallel processing. 
IMC architectures leverage device-level properties to perform in-situ Vector-Matrix Multiplication (VMM) and Boolean operations (e.g., \texttt{AND}, \texttt{NOR}, \texttt{XOR}), enabling efficient acceleration of different algorithms~\cite{roy2020memory, reis2018computing, sebastian2020memory}. IMC designs can be broadly grouped into three forms: general-purpose IMC arrays, crossbar-based VMM engines, and Content-Addressable Memories (CAMs) for parallel associative search~\cite{chi2016prime, reis2019design, moon2024afecam}. HDC aligns well with IMC due to its binary/bipolar HV and similarity computations, Static Random Access Memory (SRAM), Resistive Random Access Memory (ReRAM), Ferroelectric Field Effect Transistor (FeFET), Phase Change Memory (PCM), or Dynamic RAM (DRAM)-based HDC accelerators~\cite{liu2024reram, moon2025rex, eggimann20215, wang2024computing, mamdouh2025hide, huang2023fefet}.

Earlier work such as FSL-HD~\cite{xu2023fsl} uses a Convolutional Neural Network (CNN) to extract image embeddings before applying sequential HDC encoding and associative search in ReRAM, resulting in multi-cycle inference. ReHDC~\cite{liu2024reram} integrates ReRAM-based analog and digital crossbars but with considerable area overhead. Our design instead employs VMM in IMC for projection-based encoding and pipelines it with classification, enabling single-cycle, energy-efficient inference. 
% Prior HDC–IMC systems have not fully utilized LD sequence–based encoding or leveraged more raw input information. We address this by integrating LD-guided HDC encoding directly into IMC hardware to enhance feature extraction and overall system performance. 
\textcolor{black}{A comparative summary of existing HDC accelerators and the proposed~\Design~framework is presented in Table~\ref{tab:related_comparison}. As shown, prior approaches predominantly rely on pseudo-random hypervectors and symbolic binding operations, which increase dimensionality and hardware overhead. In contrast,~\Design~employs deterministic LD-guided encoding and enables a fully binary end-to-end in-memory inference pipeline. Unlike prior LD-based designs~\cite{moon2025rex, aygun2024sobol}, which typically use sequences mainly as quasi-random generators,~\Design~embeds LD structure directly into the projection-based HDC encoding itself, enabling a hardware-aligned representation, mitigating costly symbolic operations.}

\begin{algorithm}[t]
\caption{\Design: Learnable Deterministic HDC}
\footnotesize
\label{alg:delhd}
\begin{algorithmic}[1]
\REQUIRE Training samples $\{(\mathbf{x}_i,y_i)\}_{i=1}^N$, 
query $\mathbf{x}_q$, number of classes $K$, epochs $E$, learning rate $\eta$, 
feature dimension $F$, projection dimension D

\STATE \textbf{Sobol Projection:} 
$\mathbf{S} \gets \Phi_{\text{Sobol}}(D,F)$ 
   \STATE  $\mathbf{P} \gets \text{sign}(\mathbf{S}-0.5) \;\in \{-1,+1\}^{D \times F}$
\STATE \textbf{Training:} 
\FOR{$i=1 \dots N$}
    \STATE $\mathbf{h}_i \gets \mathbf{P}\mathbf{x}_i$
    \STATE $\mathbf{C}[y_i] \gets \mathbf{C}[y_i] + \mathbf{h}_i$
\ENDFOR
\STATE Normalize: $\mathbf{C}_k \gets \frac{\mathbf{C}_k}{\|\mathbf{C}_k\|_2}$

\FOR{$e=1 \dots E$}
    \STATE $\mathbf{H} \gets \mathbf{X}\mathbf{P}^\top$
    \STATE $Z \gets \cos(\mathbf{H},\mathbf{C})$
    \STATE $\mathcal{L} \gets \text{CE}(Z,\mathbf{y})$
    \STATE $\mathbf{C} \gets \mathbf{C}-\eta \nabla_{\mathbf{C}}\mathcal{L}$
\ENDFOR

\STATE \textbf{Inference:} $\hat{y} = \arg\max_k \cos(\mathbf{P}\mathbf{x}_q,\mathbf{C}_k)$
\end{algorithmic}

\end{algorithm}
% \vspace{-5pt}

%% file: 03_Methodology.tex
\begin{figure*}[tbp]
% \vspace{-10pt}

\centerline{\includegraphics[width=0.9\linewidth]{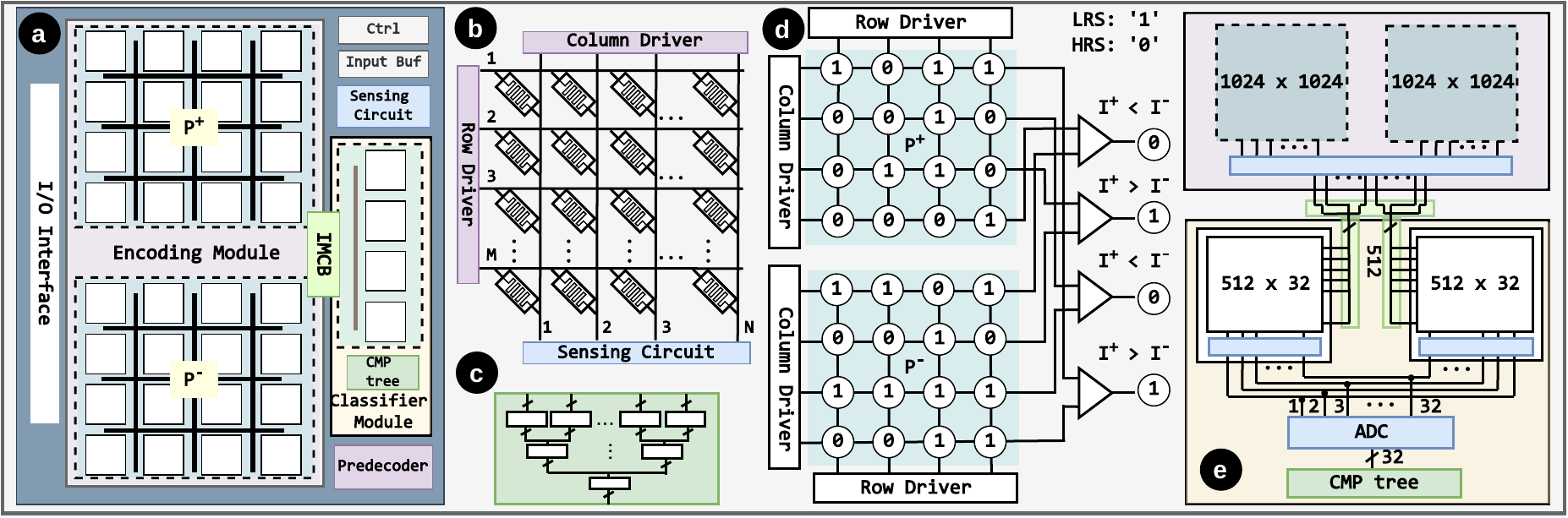}}
\vspace{-5pt}
\caption{Overview of the proposed ReRAM-based~\Design~hardware. (a) System-level organization consisting of encoding module with positive ($P^+$) and negative ($P^-$) crossbars, classifier module, and peripheral circuits (predecoder, sensing, input buffer, and control), (b) 1R ReRAM based crossbar: the core memory unit of~\Design, (c) Comparator tree for class prediction, (d) Proposed in-situ binarization technique at the sensing node, (e) Multi-crossbar organization showing data transfer between the Encoding and the Classifier Module.}
\Description{Overview diagram of the proposed ReRAM-based hardware system, showing modules and data flow.}
\label{fig:DeLHD}
\vspace{-10pt}
\end{figure*}

\section{~\Design~Framework}
\label{sec:algo}

% To unify deterministic encoding with learnable adaptation, Algorithm~\ref{alg:delhd} augments the~\Design~framework with trainable elements in both encoding and generation, yet retains the structure of a conventional HDC workflow. 
\textcolor{black}{To combine deterministic encoding with learnable adaptation, Algorithm~\ref{alg:delhd} augments the~\Design~ as an HDC framework in which deterministic projection-based encoding and similarity-based classification naturally map onto IMC primitives, enabling a low-latency and energy-efficient inference pipeline.} The overall procedure consists of three phases: (a) \textit{Deterministic HV Encoding}, (b) \textit{Learnable Class Prototypes}, and (c) \textit{Inference}.

\subsection{Learning via Deterministic HV Encoding}
\label{sec:encode}
Most existing HDC methods rely on random projection matrices derived from pseudo-random HVs or \emph{Johnson--Lindenstrauss (JL) lemma} projections~\cite{verges2025classification}. 
However, they have two main drawbacks. First, the projection matrix is \textit{stochastic}, so the encoding is not reproducible across different runs. Second, the irregular distribution of pseudo-random vectors often requires larger dimensionality to maintain separability. To address these limitations,~\Design~introduces a deterministic projection mechanism based on Sobol LD sequences. Instead of sampling a random matrix, we incorporate a fixed projection matrix,  $\mathbf{P}:= \Phi_{\text{Sobol}}(D,F) \in \mathbb{R}^{D \times F}$, 
% generated from well-known LD Sobol sequence used to 
covering high-dimensional spaces evenly~\cite{Mehran-No-Mult2023}, and further thresholded at $0.5$ to yield a deterministic bipolar encoding in $\{\pm1\}$. 

In \Design, we encode an input feature vector $\mathbf{x} \in \mathbb{R}^F$ using a simple linear mapping $\mathbf{h} = \mathbf{P}\mathbf{x}$; which is the extended forward learning trajectory of the model. 
This provides fast convergence and quasi-orthogonality by distributing features uniformly and reducing inter-dimensional correlations, resulting in more stable HVs and better class separability. Being purely linear, the encoding reduces to a single VMM. While this requires $O(D \cdot F)$ multiply–accumulate computations, software, or on conventional digital hardware platforms, e.g., Application-Specific Integrated Circuit (ASIC) or Field-Programmable Gate Array (FPGA), crossbar arrays exploit massive parallelism in-memory, effectively reducing it to $O(1)$ computation, making our approach a natural fit for IMC.

\subsection{Learnable Class Prototypes}
After encoding,~\Design~learns a set of trainable class prototypes that represent each class in D space. Each class $k \in \{1,\dots,K\}$ is assigned  a prototype vector $\mathbf{c}_k \in \mathbb{R}^D$, initialized  by accumulating encoded HVs from samples of the same class 
$\mathbf{c}_{y_i} \gets \mathbf{c}_{y_i} + \mathbf{h}_i$.
This creates centroids that summarize the distribution of samples for each class and forms the initial prototype matrix 
$\mathbf{C} = [\mathbf{c}_1;\dots;\mathbf{c}_K] \in \mathbb{R}^{K \times D}$. During training, these prototypes are iteratively refined to maximize class separability. 
% Each minibatch of encoded HVs is compared against the prototypes using cosine similarity: 
For each minibatch, encoded HVs are compared to $\mathbf{C}$ using cosine similarity,
% \vspace{-5pt}
\begin{equation}
z_{ik} = \frac{\mathbf{h}_i \cdot \mathbf{c}_k}{\|\mathbf{h}_i\| \, \|\mathbf{c}_k\|}.
\end{equation}
The similarity scores are normalized with a softmax function to produce class probabilities and
% \begin{equation}
% p_{ik} = \frac{e^{z_{ik}}}{\sum_{j=1}^{K} e^{z_{ij}}}.
% \end{equation}
the model is trained using cross-entropy loss $\mathcal{L} = -\sum_i \log p_{i y_i}.$ 
% By minimizing this loss, 
Then the prototypes are updated through gradient-based optimization. Over training epochs, this process pulls prototypes towards to samples of their own class and pushes away from others. Unlike conventional HDC that passively aggregates samples,~\Design~actively learns class representations, making the prototypes discriminative.

\subsection{Inference}
At inference time, a query sample $\mathbf{x}_q$ is deterministically projected into the D space using the same Sobol projection $\mathbf{h}_q = \mathbf{P}\mathbf{x}_q$. The encoded query is then compared with the learned prototypes. Classification 
is performed by selecting the class with the highest cosine similarity $\hat{y} = \arg\max_k \cos(\mathbf{h}_q,\mathbf{c}_k).$ 
This pipeline is highly efficient, requiring only a single VMM for encoding and a small set of similarity computations. Our proposed approach ensures consistent inference across runs, while the learned prototypes provide sharper decision boundaries than traditional HDC.

%% file: 04_Hardware.tex
\section{\Design~Architecture}
\label{sec:hw}

\textcolor{black}{\Design~adopts a ReRAM-based IMC architecture, eliminating data transfer bottlenecks and achieving low energy (sub-$\mu J$) and single-cycle inference. By exploiting the Ohmic accumulation and bitline sensing properties of ReRAM crossbars, our architecture directly executes the two core~\Design~primitives, e.g., \textit{Sobol projection} and \textit{classification}, within the memory array.}

\subsection{Hardware Organization}
\label{crossbar}
Our IMC hardware design (Fig.~\ref{fig:DeLHD}(a)) comprises two key modules: a.~encoding and b.~classifier, each implemented using ReRAM-based crossbar arrays (Fig.~\ref{fig:DeLHD}(b)). Its modules are as follows:

\textit{Encoding Module}: The encoding module consists of two ReRAM-based crossbars, marked as $P^+$ and $P^-$ in Fig.~\ref{fig:DeLHD}(a). The bipolar information is critical in HDC as it pushes HVs farther apart in the D vector space, making classes more distinguishable. We preserve the full bipolar semantics by encoding $-1$ as $(1,0)$ and $+1$ as $(0,1)$. Practically, this is realized through a dual-crossbar organization: the most-significant bit (MSB) is stored in the $P^{-}$ crossbar, while the least-significant bit (LSB) is stored in the $P^{+}$ crossbar. All cells are initially programmed to a high-resistance state (HRS), and only the cells corresponding to active components are switched to a low-resistance state (LRS). For example, a $-1$ ($+1$) element sets the corresponding $P^{+}$ ($P^{-}$) cell to LRS while the complementary cell remains in HRS.

\textit{Classifier Module}: 
On the classification side, the trained class HVs are stored in a dedicated $1024\times32$ crossbar, implemented from $32\times32$ subarrays. This layout provides sufficient width to store all class HVs while keeping the bitline length short, minimizing parasitic resistance and capacitance. The compact subarray organization also ensures better current uniformity and sensing accuracy during classification operations.

\textit{Peripheral Circuitry}: 
Our peripheral architecture incorporates hardware elements from~\cite{moon2025rex}, including shared row/column drivers and a lightweight I/O interface. Each tile uses compact current mirrors for efficient inter-tile accumulation, while the encoding module employs a current comparator~\cite{tang2009high} for one-shot current sensing. The Inter-Module Communication Bus (IMCB) establishes direct communication between the Encoding module and the Classifier module. The classifier module uses a hierarchical comparator tree (CMP\_tree in Fig.~\ref{fig:DeLHD}(c)) to determine the predicted class.

\begin{figure}[t]
    \centering
    
    \includegraphics[width=0.95\linewidth]{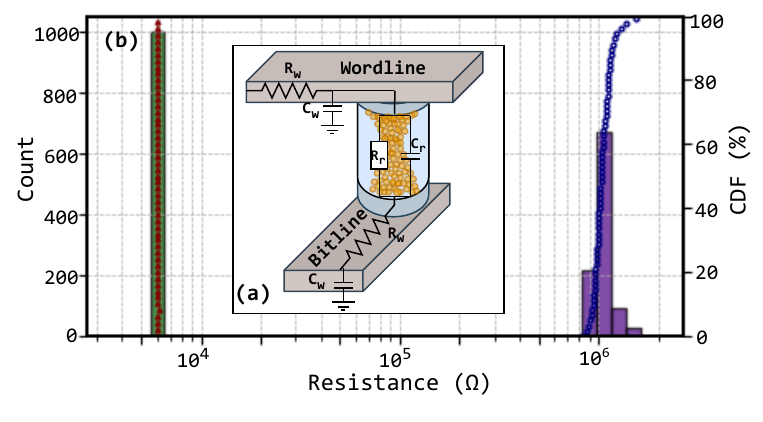}
    \vspace{-15pt}
    \caption{(a) Equivalent RC model of a ReRAM cell, including device resistance/capacitance ($R_r/C_r$) and wire RC parasitics ($R_w/C_w$). (b) Measured SET/RESET resistance distributions illustrating device-level variability.}
    \label{fig:mc}
    \Description{RC model of cell}
   \vspace{-18pt}
\end{figure}

\paragraph{Crossbar Non-idealities} 
We model our IMC engine using the Stanford ReRAM compact model~\cite{jiang2016compact}, explicitly incorporating both device behavior (cell resistance/capacitance) and wire parasitics (Fig.~\ref{fig:mc}(a)). This model captures key non-idealities such as spatial variability, cycle-to-cycle noise, nonlinear switching, IR drop, and read noise, as illustrated in Fig.~\ref{fig:mc}(b), and enables~\Design~to operate robustly under these effects. To ensure robust operation, we tile the logical $1024 \times 1024$ array into sixteen $64 \times 128$ subarrays. Limiting the bitline height to 64 cells keeps IR-drop and line-resistance ($0.76\Omega$/cell)~\cite{knudsen2008nangate} effects small, while the wider 128-column layout preserves high parallelism. This tiled organization maintains signal integrity during large-scale VMM and avoids the scalability issues associated with a single large crossbar. 

The SET/RESET distributions in Fig.~\ref{fig:mc}(b) show roughly two orders of magnitude separation between LRS (K$\Omega$) and HRS (M$\Omega$). In a $64\times 128$ crossbar, this inherently limits worst-case leakage relative to the selected conduction path. Moreover, sneak-path currents plays a negligible role during VMM, since all rows are fully driven with respective $V_{read}$, leaving no half-selected devices to form unintended paths~\cite{chen2018neurosim}. Since Sobol projection matrix and class HVs are written once, a simple 1R crosspoint array is sufficient, and efficient write schemes such as SWIPE~\cite{gonugondla2020swipe} can be used to further minimize programming noise. These choices make~\Design~well suited for robust, low-energy, fully parallel IMC operation.

 \begin{figure*}[t]
    \centering
    % \vspace{-15pt}
    \includegraphics[width=0.91\linewidth]{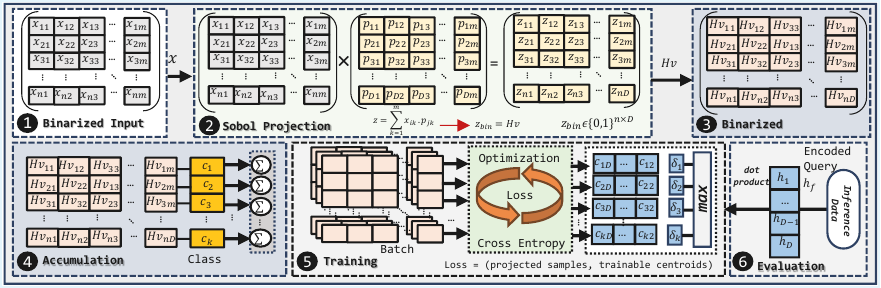}
      \vspace{-5pt}
     \caption{
    The proposed~\Design~training and evaluation framework.
\textcircled{1}~\textbf{Binarized Input:} Raw features $X \in \mathbb{R}^{n \times m}$ are thresholded to binary form.
\textcircled{2}~\textbf{Sobol Projection:} Inputs are mapped to a high-dimensional space using a Sobol quasi-random projection matrix $P \in \mathbb{R}^{D \times m}$, producing real-valued HVs $Z$.
\textcircled{3}~\textbf{Binarization:} Projected HVs are thresholded to obtain $Z_{\text{bin}} \in \{0,1\}^{n \times D}$.
\textcircled{4}~\textbf{Accumulation:} Class-wise aggregation initializes centroids $C \in \mathbb{R}^{k \times D}$.
\textcircled{5}~\textbf{Training:} Centroids are refined via cross-entropy optimization to enhance class separation.
\textcircled{6}~\textbf{Evaluation:} At inference, a query HV is compared with trained centroids using dot-product similarity, and the class with the highest score is chosen.}
    \label{fig:soft_pro}
    \Description{A six-stage workflow diagram showing the XL-HD process: binarized input, Sobol projection, hypervector binarization, class-wise centroid accumulation, cross-entropy-based training, and dot-product-based evaluation.}
      \vspace{-10pt}
\end{figure*}

\subsection{End-to-End Algorithm/Hardware Mapping}
% We further co-design across the algorithm and hardware, herein referred to as~\HDesign, to reduce the computational complexity of prior methods~\cite{imani2019binary, rahimi2016robust,  liu2024reram}. 
\textcolor{black}{Traditional HDC relies on binding, bundling, and multi-bit quantization, all of which require repeated logic operations, large memory accesses, and increased storage per dimension~\cite{kanerva2009hyperdimensional, hernandez2021onlinehd}. In contrast, we co-design across the algorithm and hardware, herein referred to as~\HDesign, to enforce a fully binary representation throughout the inference pipeline to minimize hardware cost while maintaining high accuracy.}

\textcolor{black}{Fig.~\ref{fig:soft_pro} presents~\HDesign, which is designed to align algorithmic steps with efficient hardware execution. The pipeline starts with input binarization in \textcircled{1}, where raw feature vectors $X \in \mathbb{R}^{m \times F}$ are thresholded to produce binary activations. In \textcircled{2}, these binarized inputs are projected into a high-dimensional space using the Sobol-based projection matrix $P \in \mathbb{R}^{D \times F}$. In \textcircled{3}, the projected values are converted into binary hypervectors through \textit{in-situ} binarization. The projection matrix $P$ is decomposed into two binary masks and mapped onto the $P^{+}$ and $P^{-}$ crossbars, where the binary input features from \textcircled{1} drive the wordlines with a small read voltage $V_{read}$, generating differential bitline currents $(I_d^{+}, I_d^{-})$. A compact current comparator performs the binarization at \textcircled{3} directly in memory:}
\textcolor{black}{
\begin{equation}
h_d =
\begin{cases}
1, & I_d^{+} > I_d^{-} \\
0, & \text{otherwise}
\end{cases}
\end{equation}}

\textcolor{black}{This mechanism generates the encoded binary hypervector without requiring ADCs, DACs, or floating-point computation.}

\textcolor{black}{During training, \textcircled{4} accumulates encoded hypervectors belonging to the same class to form class-wise centroids. In \textcircled{5}, these centroids are refined using cross-entropy optimization. These training steps are performed offline in software to preserve learning flexibility. Once training converges, the resulting class hypervectors are binarized and programmed into the classifier crossbar arrays.}

\textcolor{black}{At inference time, the entire pipeline executes on the IMC hardware. In \textcircled{6}, a query hypervector $h_q \in \{0,1\}^{D}$ drives the wordlines of the classifier crossbar while the trained class hypervectors $c_k$ are stored along the bitlines. The resulting currents implement binary dot-product similarity:}
% \vspace{-10pt}
\textcolor{black}{\begin{equation}
z_k = h_q \cdot c_k = \sum_{i=1}^{D} h_{q,i} c_{k,i}.
\end{equation}}

\textcolor{black}{A hierarchical comparator tree identifies the class with the highest similarity score. By mapping each stage of the algorithm to efficient IMC primitives,~\Design~enables a fully integrated inference pipeline while preserving flexibility for software-based training.}

%% file: 05_Experimental_Results.tex
\section{Evaluation}
\label{evaluation}

\begin{table}[t]
\centering
% \footnotesize   % reduced from \small
\caption{Performance comparison of HDC models across datasets ($D = 10K$) [Accuracy in \%]}
\label{tab:hdc_models}
\vspace{-10pt}
\renewcommand{\arraystretch}{0.8}
\setlength{\tabcolsep}{10pt}   % tightened to fit column

\begin{tabular}{lccc}
\toprule
\textbf{Model} & \textbf{MNIST} & \textbf{UCIHAR} & \textbf{ISOLET} \\
\midrule
HDC                         & 80.36 & 82.46 & 87.42  \\
SearcHD~\cite{imani2019searchd}  & 84.43 & 82.31 & 83.47 \\
QuantHD~\cite{imani2019quanthd}  & 89.28 & 91.25 & 92.70 \\
OnlineHD~\cite{hernandez2021onlinehd} & 96.00 & 95.00 & 96.00 \\
LeHDC~\cite{duan2022lehdc}        & 94.70 & 95.20 & 94.90   \\
BinHD~\cite{imani2019binary}      & ---   & 95.70 & 91.00 \\
XL-HD                            & 93.49 & 95.22 & 94.87  \\
H-XL-HD                 & 93.20 & 91.36 & 91.00  \\
\bottomrule
\end{tabular}

\vspace{2pt}
\begin{flushleft}
\tiny\emph{GPU: NVIDIA GTX 3050, CPU: Intel(R) Core(TM) i7, 2.30GHz.}
\end{flushleft}
\vspace{-15pt}
\end{table}

\begin{table}[t]
\centering
% \footnotesize
\caption{XL-HD Specification}
\label{tab:components}
\vspace{-10pt}
\renewcommand{\arraystretch}{0.8}
\setlength{\tabcolsep}{5pt}

\begin{tabular}{lccc}
\toprule
\textbf{Component} & \textbf{Area ($\mu m^2$)} & \textbf{Power (mW)} & \textbf{Size} \\
\midrule
XBar\_P      & 189213.64 & 245 & 1024$\times$1024 \\
XBar\_C      & 13290.18  & 3.98 & 1024$\times$32   \\
Data Buffer & 1089.54  & 0.13 & 1$\times$1024    \\
CMP tree     & 1081.82  & 0.62 & 32:1             \\
ADC          & 28.69    & 2.47 & 8-bit            \\
IMCB         & 327.68   & 2.09 & 512-bit width    \\
\bottomrule
\end{tabular}

\vspace{2pt}
\begin{flushleft}
\tiny\emph{Single-level-cell: $R_{on}=6\,\text{K}\Omega$, $R_{off}=1\,\text{M}\Omega$ @ $V_{set}= -2$V, $V_{reset}= 2.8$V  }
\end{flushleft}
\vspace{-20pt}
\end{table}

In this section, we evaluate the performance of the proposed \Design\ framework on three benchmarking datasets, i.e., MNIST~\cite{MNIST}, ISOLET~\cite{frank2010uci}, and UCIHAR~\cite{anguita2012human}, 
% and FACE~\cite{angelova2005pruning}, 
covering a wide range of applications like computer vision, voice-, and activity-
% , and face 
recognition. \textcolor{black}{We compare~\Design~against several state-of-the-art HDC models and analyze both software execution and the proposed ReRAM-based IMC implementation. Our evaluation demonstrates that~\Design~achieves competitive accuracy while significantly reducing inference latency and energy through its deterministic projection pipeline and an inference path executed entirely within the IMC architecture.}

\subsection{Experimental Setup}

At the software level,~\Design~was implemented in Python 3.1 using PyTorch 2.1.0 and executed on an NVIDIA RTX 3050 GPU with an Intel Core i7 CPU (2.30\,GHz). To ensure a fair comparison across HDC models, we evaluate multiple HV dimensionalities, since performance in HDC is strongly dependent on the choice of $D$.

\textcolor{black}{At the hardware level, we base our ReRAM device modeling on the Stanford physics-based Verilog-A compact model, which provides an experimentally calibrated parameter-extraction methodology and explicitly validates switching/resistance distributions against fabricated TiN/TiO$_x$/HfO$_x$/Pt planar ReRAM devices~\cite{jiang2016compact}}. We estimated the ReRAM crossbar area assuming a standard $4F^2$ 1R cell structure, and obtain power directly from Cadence Spectre simulations.
 
We synthesized the CMP tree, input buffer, controller in Cadence Genus with the NanGate 45-nm open-cell library~\cite{knudsen2008nangate}, and modeled IMCB using CACTI7.0~\cite{balasubramonian2017cacti}. We incorporate current comparator proposed in~\cite{tang2009high}, and the crossbar peripheral units as detailed in~\cite{chen2018neurosim}.
 % \vspace{-5pt}
\subsection{\Design~Accuracy}

Table~\ref{tab:hdc_models} compares the classification accuracy of~\Design~and~\HDesign~against several representative HDC models. Both our approaches consistently outperform conventional, multi-model, and retraining-based HDC implementations across all datasets. 
% Along with baseline models, for a fair comparison, we furthermore analyze the performance of our model with state-of-the-art HDC models. 
While OnlineHD~\cite{hernandez2021onlinehd}, under fixed learning rate, achieves competitive accuracy for MNIST and ISOLET,~\Design~achieves superior performance on UCIHAR improving accuracy by $0.22$\%. 
% and $2$\%, respectively.
~\Design~maintains iso-accuracy with LeHDC~\cite{duan2022lehdc}. Comparing the fully binary implementation, BinHD~\cite{imani2019binary} to our~\HDesign, BinHD achieves a $4.34$\% higher accuracy on UCIHAR; however, \HDesign~
% achieves a $1.70$\% improvement on FACE and 
maintains iso-accuracy on ISOLET, proving our approach  to be robust across diverse tasks. 

% \vspace{-5pt}
\subsection{\Design~Efficiency}

\begin{table}[t]
\centering
% \footnotesize
\caption{Per-sample energy (µJ): CPU vs.~\Design~for inference at $D = 1\text{K}$}
\label{tab:energy_breakdown}
\vspace{-10pt}
\renewcommand{\arraystretch}{0.8}
\setlength{\tabcolsep}{4.5pt}

\begin{tabular}{lcccccc}
\toprule
\multirow{2}{*}{\textbf{Dataset}} 
& \multicolumn{2}{c}{\textbf{Encoding}} 
& \multicolumn{2}{c}{\textbf{Classification}} 
& \multicolumn{2}{c}{\textbf{Total}} \\
\cmidrule(lr){2-3} \cmidrule(lr){4-5} \cmidrule(lr){6-7}
 & CPU & \Design & CPU & \Design & CPU & \Design \\
\midrule
MNIST  & 95.55  & 0.384 & 7.35   & 0.0061 & 102.90  & 0.391 \\
UCIHAR & 49.23  & 0.275 & 11.36  & 0.0037 & 60.59   & 0.279 \\
ISOLET & 63.36  & 0.302 & 29.24  & 0.0159 & 92.60   & 0.319 \\
% FACE   & 185.00 & 0.298 & 832.50 & 0.0012 & 1017.50 & 0.299 \\
\bottomrule
\end{tabular}
\vspace{-6pt}
\end{table}

\begin{figure}[t]
    \centering
    \includegraphics[width=1.02\linewidth]{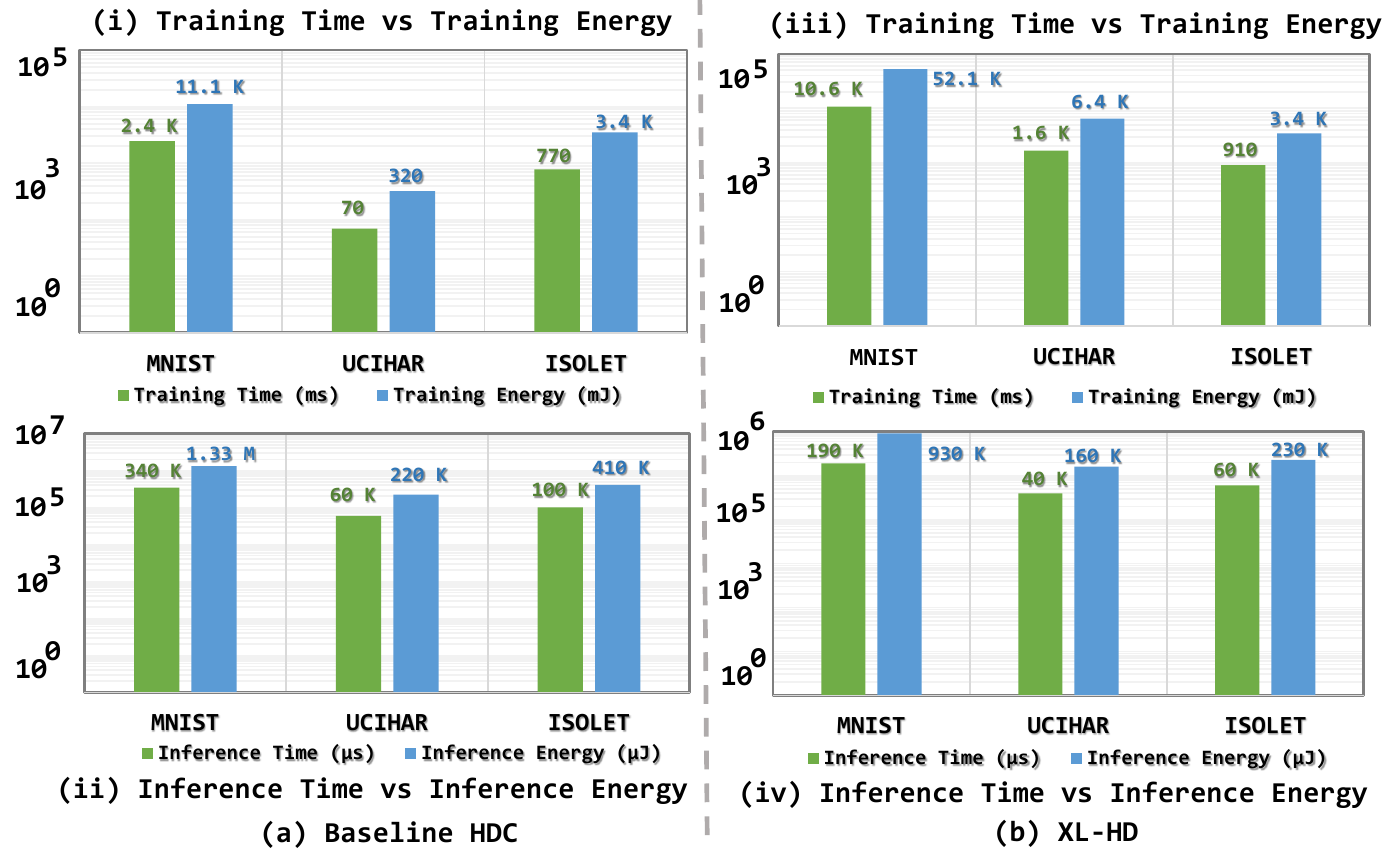}
\vspace{-20pt}
    % \vspace{-20pt}
    \caption{Training and inference time and energy comparison between Baseline HDC and~\Design~across three datasets with $D = 10K$.}
    \label{fig:energy_runtime}
\vspace{-15pt}
\end{figure}
Table~\ref{tab:components} summarizes the area and power breakdown of the core components in the~\Design, which achieves single-cycle inference in a compact $0.395  mm^2$ footprint. For fair comparison, HyperMetric~\cite{xu2025hypermetric} is scaled from 40nm to 45nm using DeepScaleTool~\cite{sarangi2021deepscaletool}; even though at $D=1024$, it occupies $0.134 mm^2$, it needs $1568$ cycles for one classification. 

\textcolor{black}{A recent ReRAM-based HDC design~\cite{liu2024reram} employs a hybrid Analog and Digital Processing-in-Memory (APIM \& DPIM) architecture, where DPIM arrays implement XOR operations and APIM arrays perform accumulation. This dual-fabric design introduces significant hardware overhead, with the DPIM block alone occupying $2\times$ the area of the APIM. In contrast,~\Design~reformulates encoding as a projection that maps directly to crossbar VMM, enabling both encoding and classification within a single ReRAM crossbar fabric. ReX-HD~\cite{moon2025rex} reports a total footprint of $0.194\,\text{mm}^2$, power consumption of $69.65\,\text{mW}$, and inference energy of approximately $1.09\,\mu\text{J}$/sample. In comparison,~\Design~achieves inference energy below $0.40\,\mu\text{J}$/sample at $D=1024$, corresponding to $\sim2.7\times$ lower inference energy and $\sim5.5\times$ lower power consumption.}

Table~\ref{tab:energy_breakdown} reports per-sample inference energy across datasets for CPU vs IMC implementation. Encoding dominates the total energy ($\approx60$--$70$\%), while classification scales with the number of classes (e.g., $3.7 nJ$ for UCIHAR vs. $15.9 nJ$ for ISOLET). Inference energy remains below $0.40 \mu J$, demonstrating the efficiency of our in-memory design. Overall,~\Design~reduces inference energy by $2-4$ orders of magnitude, yielding an average efficiency gain of $99.7\%$ over CPU execution. 
% We also present a GPU-based evaluation of~\Design~in 
Fig.~\ref{fig:energy_runtime} further compares GPU execution of~\Design~against baseline HDC~\cite{imani2019binary} at $D = 10K$. \textcolor{black}{While~\Design~incurs a moderate increase in training overhead due to gradient-based optimization, it significantly reduces inference latency and energy through its deterministic projection and lightweight classification pipeline.}

\subsection{~\Design~Execution Time}

We analyze the execution time of~\Design~w.r.t. prior implementations. ReX-HD~\cite{moon2025rex} incurs a significant latency by reloading feature HVs during every inference, scaling with the number of input features. 
% Our approach avoids this bottleneck by programming the bipolar Sobol projection matrix once at deployment, enabling efficient and low-latency inference thereafter.
\textcolor{black}{In contrast,~\HDesign~targets inference-only hardware, with training performed offline in software. The bipolar Sobol projection matrix and trained binary class HVs are programmed once at deployment, eliminating repeated memory transfers. As a result, end-to-end inference completes in two evaluation cycles, achieving a steady-state throughput of one inference per cycle through pipelining.} On ASICs, tinyHD~\cite{khaleghi2021tiny} offers a compact footprint, but suffers from  $12.2k$ cycles encoding latency. Later, GENERIC~\cite{khaleghi2022generic} improves this but still remains $256\times$ slower than~\Design.

\subsection{~\Design~Robustness}

\begin{figure}[t]
    \centering
    % \vspace{-15pt}
    \includegraphics[width=0.95\linewidth]{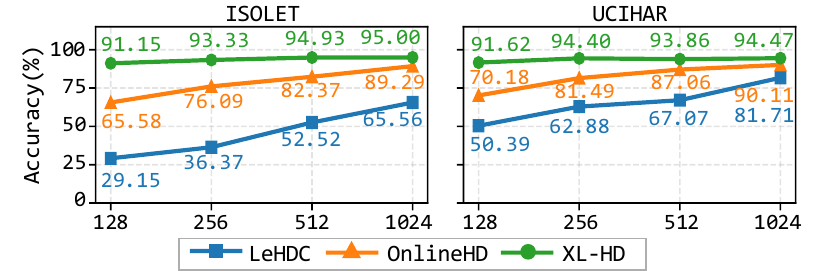}
    \vspace{-10pt}
    \caption{Accuracy comparison at lower HV dimensions.}
    \label{fig:acc_dim}
    \vspace{-10pt}
   % \vspace{-15pt}
\end{figure}

Fig.~\ref{fig:acc_dim} highlights a dimension-accuracy trade-off in the low-dimensional regime, comparing~\Design~to the strongest models. Across all tested dimensions,~\Design~consistently outperforms, maintaining high accuracy even under aggressive dimensionality reduction. In contrast, OnlineHD~\cite{hernandez2021onlinehd} requires $4-8$K dimensions to perform well with 1-bit hypervectors, and LeHDC~\cite{duan2022lehdc} achieves strong accuracy only at high dimensions.

Notably,~\Design~shows $39.04$\% and $12.13$\% higher accuracy at $D=128$ than LeHDC at $D=1K$ for the ISOLET and UCIHAR, respectively. 

We further evaluate our design under a bit-flip fault model
% , where each bit is flipped independently with a probability ranging from 
($0\%$ to $15\%$ Bit Error Rate (BER)), and Table~\ref{tab:noise_quality_loss} shows that~\Design~exhibits 
% significantly greater robustness to hardware faults compared to 
higher resilience than deep neural network (DNN) and OnlineHD~\cite{hernandez2021onlinehd}, with minimal degradation even at high BER. Its resilience remains strong even with $10\times$ reduced dimensionality, highlighting its suitability for fault-prone edge environments.

\begin{table}[t]
\centering
% \footnotesize
\caption{Quality loss (\%) under bit error rates (BER).}
\label{tab:noise_quality_loss}
\vspace{-10pt}
\renewcommand{\arraystretch}{0.8}
\setlength{\tabcolsep}{4.5pt}

\begin{tabular}{lccccc}
\toprule
\textbf{Hardware Error (BER)} & \textbf{1\%} & \textbf{2\%} & \textbf{5\%} & \textbf{10\%} & \textbf{15\%} \\
\midrule
DNN                     & 3.9  & 9.4  & 16.3 & 26.4 & 40.0 \\
OnlineHD~\cite{hernandez2021onlinehd} & 0.0  & 0.0  & 0.90 & 3.10 & 5.20 \\
\Design\,(D=10K)        & 0.0  & 0.0  & 0.18 & 1.07 & 3.34 \\
\Design\,(D=1K)         & 0.08 & 0.11 & 0.49 & 2.33 & 5.16 \\
\bottomrule
\end{tabular}
\vspace{-12pt}
% \vspace{-20pt}
\end{table}

%% file: 06_Conclusion.tex
\section{Conclusion}
In this paper, we present~\Design, a deterministic, hardware-friendly HDC framework co-designed with ReRAM-based IMC architectures. By replacing traditional binding and quantization with Sobol-based projections and a fully binary pipeline,~\Design~removes major overheads and enables direct in-situ crossbar execution. 
% It maintains high accuracy across datasets while significantly lowering area, latency, and energy. 
Our results show that ~\Design~ach-ieves competitive accuracy with a $0.395 mm^2$ footprint, $0.40$$\mu J$ inference energy, and single-cycle throughput, combining HDC accuracy with IMC efficiency. We observe only $3.34$\% and $5.16$\% degradation at $BER = 15$\% for~\Design~with $D = 10K$ and $D = 1K$, respectively, demonstrating strong resilience under high hardware fault rates.

% \footnote{Code is available at: 
% \href{https://github.com/ACEDLab/XL-HD}{%
% \raisebox{-0.25em}{\includegraphics[height=1em]{figures/github.png}}%
% }
% \ \href{https://github.com/ACEDLab/XL-HD}{\url{https://github.com/ACEDLab/XL-HD}}
% }

%% file: 07_Ack.tex
\section*{Acknowledgments}
This work is supported by NASA grant 80NSSC25C0335, Vernon $\&$ Ruby Langlinais Non-Endowed Research Fund, Lockheed Martin Corporation Endowed Professorship Fund, an award from NASA, and gifts from NVIDIA and Google.